\documentclass[longbibliography,aps,prl,twocolumn,superscriptaddress,groupedaddress]{revtex4}
\usepackage{graphicx}  
\usepackage{dcolumn}   
\usepackage{bm}        
\usepackage{amssymb}   
\usepackage{physics} 
\usepackage[dvipsnames]{xcolor} 
\usepackage{amsmath}
\usepackage{hyperref}
\usepackage{color,newfloat}
\usepackage{float}
\hypersetup{
    colorlinks=true,
    linkcolor=Blue,
    citecolor=Blue,
    urlcolor=Blue
}
\usepackage[caption = false]{subfig}
\usepackage{lineno}

\begin{document}

\title{High fidelity photon-photon gates by scattering off a two-level quantum emitter}
\author{Love A. Pettersson}
\email{love.pettersson@nbi.ku.dk}
\affiliation{Center for Hybrid Quantum Networks (Hy-Q), Niels Bohr Institute, University of Copenhagen, Blegdamsvej 17, 2100 Copenhagen, Denmark.}
\author{Victor R. Christiansen}
\email{victorrc@phys.au.dk}
\affiliation{Center for Complex Quantum Systems, Department of Physics and Astronomy, Aarhus University, Ny Munkegade 120, DK-8000 Aarhus C, Denmark}
\author{Klaus Mølmer}
\email{klaus.molmer@nbi.ku.dk}
\affiliation{Center for Hybrid Quantum Networks (Hy-Q), Niels Bohr Institute, University of Copenhagen, Blegdamsvej 17, 2100 Copenhagen, Denmark.}
\author{Anders S. Sørensen}
\email{anders.sorensen@nbi.ku.dk}
\affiliation{Center for Hybrid Quantum Networks (Hy-Q), Niels Bohr Institute, University of Copenhagen, Blegdamsvej 17, 2100 Copenhagen, Denmark.}

\date{\today}

\begin{abstract}
We present a scheme for implementing a high-fidelity non-linear phase shift on a photonic state.
The scheme is based on repeated scattering off a two-level quantum emitter embedded in a chiral or one-sided waveguide. The waveguide is equipped with elements inducing second-order dispersion and temporal phase shifts, which effectively form a harmonic trap and confine the photon pulses to a Gaussian shape.
The same quantum emitter can be used for each scattering, and thus, only one quantum emitter is needed in this scheme.
To illustrate the application of our scheme for photonic quantum computing and quantum communication, we analyze the implementation of a control-Z gate and a deterministic Bell-state analyzer for photonic qubits. Through numerical optimization, we show that we can reach a control-Z gate fidelity of $\mathcal{F} \sim 99.2\%$ ($\mathcal{F} \sim 96\%$) and a success probability of $P_s \sim 99.6 \%$ ($P_s\sim 98 \%$) for a Bell-state measurement with $N=17$ ($N=5$) scatterings.
\end{abstract}

\maketitle
\textit{Introduction---}
Using photons as carriers of quantum information offers several significant advantages, including long coherence times due to their weak interaction with the environment, ease of manipulation, and straightforward distribution~\cite{Flamini_2019, SlussPQCRev, Wang2020}.
However, a challenge for photonic quantum information processes lies in implementing a two-qubit gate between photonic qubits, as photons do not interact directly.
There are several proposals to realize universal optical quantum computing \cite{Oneway, FBQC, NielsenOQC, QuandelaSpinOpticalComputing, löbl2023losstolerant, Knill2001}, quantum simulation~\cite{Sparrow2018}, and quantum networks~\cite{AllPhotonicRepeaters, SophiaRGS, LukinTwoWay, NicolasTwoWay}, which circumvent this problem by combining linear optics and the inherent non-linearity of quantum measurements~\cite{ScheelMBQC}.
While these schemes benefit from the ease and high fidelity with which linear optics operations and measurements can be performed, they are inherently probabilistic, which imposes a large resource overhead~\cite{BrowneDaniel, ResourceCostLOQC}.

\begin{figure}
\includegraphics[scale=0.65]{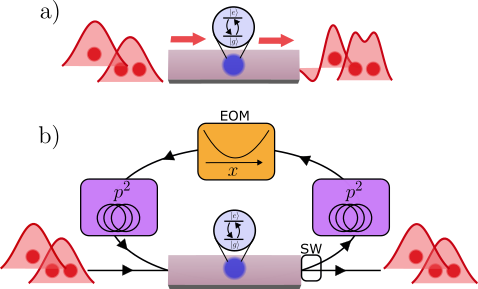}
\caption{\label{fig:intro-figure} A schematic illustration of the main components of our phase gate protocol. a) A two-photon or single-photon pulse is scattered off a two-level quantum emitter embedded in a chiral waveguide. Due to dispersion and spectral entanglement induced by the two-level system, the pulses leave the device distorted. b) Adding a harmonic trap for the light pulse through the sequential evolution of a second-order dispersive element ($p^2$) and a temporally varying phase (EOM) between the repeated scatterings mitigates the distortion effects on the input pulses. Thus, the pulses will leave the device undistorted through the switch (SW) after accumulating the desired phase.}
\end{figure}
A way to circumvent the significant resource overhead is to mediate an interaction between photonic qubits by non-linear matter~\cite{Chang2014}. 
This approach enjoys the feature of being deterministic.
A promising non-linear system is a two-level quantum emitter as it possesses an inherent non-linearity~\cite{RoyNonLinRev, SherNonLinRev} by only being able to absorb one photon at a time.
This can be utilized to mediate a photon-dependent phase shift.
While the multimode nature of photon wavepackets makes it challenging to realize such photon-dependent phase shifts with high fidelity due to wavepacket deformation~\cite{ShapiroFirstPaper, SecondShapiro} (see Fig.~\ref{fig:intro-figure} a), in later years, a plethora of schemes have been proposed to circumvent this problem~\cite{Schrinski_2022, passiveCphaseGate, FanPhotonSorter, ShapiroCascadedCZGate, CombesMathCounter, DispersionEngCZGate}.
To mention a few, these proposed schemes include sending counter-propagating wavepackets through a chiral waveguide coupled to an array of emitters~\cite{Schrinski_2022, passiveCphaseGate}, cascading the scattering of co-propagating photons and exploiting pulse reshaping~\cite{ShapiroCascadedCZGate}, or optimizing input temporal pulse shapes to realize two orthogonal output mode shapes~\cite{FanPhotonSorter}.

In this letter, we propose a protocol for a high-fidelity photon-dependent phase shift between two co-propagating photonic qubits in a chiral or one-sided waveguide~\cite{LodahlChiral, NilsChiral} coupled to a single two-level system.
The two-level system (TLS) offers extremely large nonlinearity and thus presents the simplest system for achieving non-linear phase shifts on the single-photon level.
As in Ref.~\cite{ShapiroCascadedCZGate}, instead of imparting a $\pi$-nonlinear phase shift from a single round of scattering, we cascade the scattering $N$ times to build up a $\pi$ phase shift slowly.
To address wavepacket deformation and the accompanying phase corruption \cite{ShapiroCascadedCZGate}, we introduce an effective harmonic temporal trap that confines the photonic wavepacket to a Gaussian mode.
The trap is achieved by a combination of temporally varying phases and second-order dispersion, as shown in Fig.~\ref{fig:intro-figure} b).
Similar pulse trapping was also proposed in Ref.~\cite{TemporalTrappingCZgate}; however, here we leverage the strong nonlinearity of a two-level system rather than a weak bulk nonlinearity.
Furthermore, in contrast to Ref.~\cite{ShapiroCascadedCZGate}, we do not aim to reshape the pulse after each scattering but rather trap the pulse to a Gaussian mode throughout the cascading.
This lets us achieve high-fidelity two-qubit operations with orders of magnitude less scatterings than Ref.~\cite{ShapiroCascadedCZGate}.
Furthermore, our scheme does not need experimentally challenging mode-selective operations~\cite{PhysRevLett.120.213601, PhysRevX.5.041017, Eckstein:11}, and enjoys the benefit of re-using one emitter, thus not relying on an array of emitters.
We demonstrate that our setup can achieve a two-qubit control-Z gate with a fidelity of $\mathcal{F} > 99.2\%$ ($\mathcal{F} \sim 96\%$) and a Bell-state measurement with a success probability of $P_s>99.5 \%$ ($P_s \sim 98\%$) with $N=17$ ($N=5$) scatterings.
The high-fidelity control-Z gate is a key instrument for quantum simulation~\cite{KasperQuantumSim}, and the high success probability Bell-state measurement may drastically increase the performance of photonic quantum computing~\cite{FBQC, StefanoBen} and quantum communication protocols~\cite{AllPhotonicRepeaters, pettersson2025QR}.
\textit{Protocol---}
Our goal is to impart a photon number dependent phase shift such that the difference between the single-photon phase shift $\phi_L$ and the two-photon phase shift $\phi_{NL}$ amounts to $\pi$.
That is, we wish to implement the transformation
\begin{subequations}
\begin{align}
    \ket{1} &\xrightarrow[]{} e^{i\phi_L}\ket{1} \\
    \ket{2} &\xrightarrow[]{} e^{i\phi_{NL}} \ket{2}
\end{align}
\end{subequations}
with $\phi_{NL} - 2 \phi_L = \pi$, and where $\ket{1}$ ($\ket{2}$) is a single (two) photon Fock state.
Here, we consider the photons to be described by the states
\begin{subequations}
\begin{align}
    \ket{1} &= \int dk \phi(k)\hat{a}^{\dagger}(k)\ket{\O}, \\
    \ket{2} &= \frac{1}{\sqrt{2}}\int dk dp \psi(k, p) \hat{a}^{\dagger}(k) \hat{a}^{\dagger}(p)\ket{\O},
\end{align}
\end{subequations}
where $\ket{\O}$ is the multimode vacuum state, $\phi(k)$ and $\psi(k, p)=\phi(k)\phi(p)$ the mode functions of the photons, and $\hat{a}^{\dagger}(k)$ the operator associated with creating a photonic excitation with momentum $k$. 
The central component for realizing the non-linear phase shift is a two-level quantum emitter embedded in a chiral waveguide, as illustrated in Fig.~\ref {fig:intro-figure}.
Scattering a single and two-photon input state, the output state is $\int dk \Phi(k)\hat{a}^{\dagger}(k)\ket{\O}$ and $\int dk dp \Psi(k, p)\hat{a}^{\dagger}(k)\hat{a}^{\dagger}(p)\ket{\O} / \sqrt{2}$, respectively.
Here, the output single photon mode function $\Phi(k) = \mathcal{T}(k)\phi(k)$ is the input mode function transformed by a linear transmission coefficient $\mathcal{T}(k)$, and the two-photon mode function $\Psi(k, p) = \int dk' dp' \mathcal{S}(k, p;k',p')\psi(k', p')$ is the input mode function transformed by the scattering matrix $\mathcal{S}$ for a two-level emitter~\cite{ShenAndFenScattering, ShenAndFenSecond}.
As an alternative, the problem can also be viewed in the time domain as the result of integrating the master equation dynamics of a time-dependent incoming quantum pulse scattering off a two-level system.
In this work, we find the scattered quantum state using both of these methods and confirm that they yield equivalent results.
For more information on the technical details of calculating the scattering quantum state, see the supplementary material.
\begin{figure}
\includegraphics[scale=0.8]{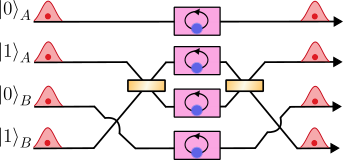}
\caption{Illustration of the circuit for implementing the control-Z gate. The circuit interferes the $\ket{11}_{AB}$ component, leading to $\ket{20}$ and $\ket{02}$ components subject to the desired phase shift, whereas the other components travel as independent photons. The subsequent beamsplitter recombines the modes and yields the control-Z gate. 
 \label{fig:CZ_gate} }
\end{figure}
Although the quantum emitter mediates a photon-number-dependent phase shift, it also distorts the incoming wavepackets.
For the single photon wavepacket, the distortion is due to a frequency-dependent phase shift.
For the two-photon wavepacket, the scattering spreads the wavepacket out in frequency, leading to spectral entanglement within the photon pair~\cite{ShenAndFenScattering, ShenAndFenSecond, passiveCphaseGate}.
Critically, the two-photon mode function is distorted differently from two independent single-photon mode functions; i.e., they become partially distinguishable, as $|\int dk dp \Phi^*(k)\Phi^*(p) \Psi(k,p)| < 1$.
This is not ideal if the non-linear phase shift is to be used for quantum information processes, as photon distinguishability leads to erroneous operations~\cite{YardPaesaniDisting}. 
To circumvent mode function distortion, we propose cascading the off-resonant scattering with reshaping of the light pulse between scatterings, leading to a loop-like evolution as illustrated in Fig.~\ref{fig:intro-figure}b).
The same way a harmonic trap confines a quantum particle to a Gaussian ground state, a second-order dispersive element and a temporally varying phase can confine an optical pulse solution to Maxwell's Equations in the time and frequency domain.
Application of the evolution operators $e^{-i\lambda_1 p^2 }$, $ e^{-i \lambda_2 x^2}$, and $ e^{-i \lambda_3 p^2}$ combine into $ \mathcal{U}_{HO}(\Delta t) = \exp\left[-i\Delta t \Omega\left(\frac{\sigma^2 p^2}{2} + \frac{x^2}{2 \sigma^2}\right)\right]$, with $\lambda_1 = \lambda_3 = \sigma^2 \tan{\left(\frac{\Omega \Delta t}{2}\right)}/2$ and $\lambda_2 = \sin{\left(\Omega \Delta t \right)} /2 \sigma^2$ (see supplementary material for derivation), with $x$ being position and $p$ momentum, and where $\sigma$ is the temporal width of a Gaussian wavepacket.
In the time domain, with $x$ and $p$ representing time and frequency variables, we can use discrete optical elements~\cite{Karpi_ski_2016} to achieve the same trapping of light pulses as for a quantum particle.
If we consider the interaction with the two-level system as a small perturbation to the harmonic oscillator, the intuition is that the pulse will mainly remain Gaussian, as it is the harmonic oscillator's ground state.
\begin{figure}
\includegraphics[scale=0.38]{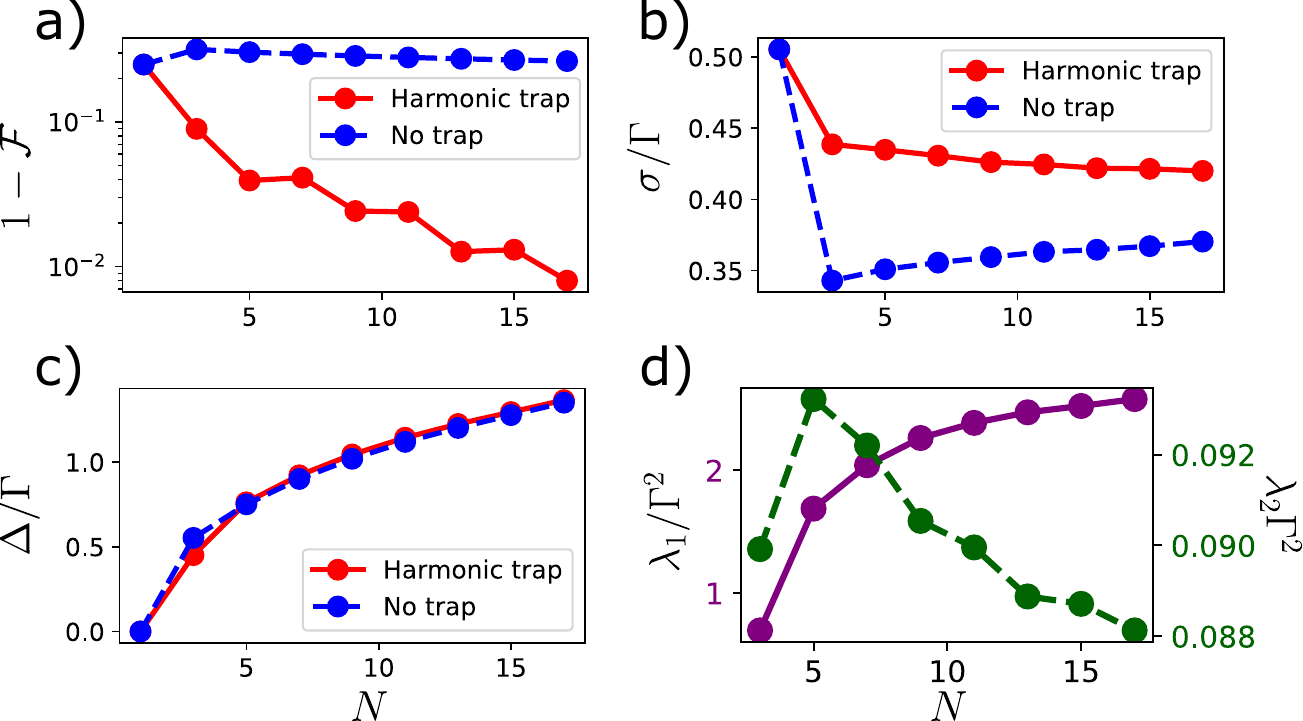}
\caption{
a) Infidelity of the gate as a function of the number of scatterings, with the harmonic trap (red) and without (blue). b) Optimal bandwidth $\sigma$ in units of the decay rate $\Gamma$ of the two-level system, and c) the optimal detuning in units of $\Gamma$. d) Optimal trapping parameters $\lambda_1$ (purple) and $\lambda_2$ (green) in units of the decay rate $\Gamma$.
 \label{fig:CZ_gate_results} }
\end{figure}
\textit{Control-Z gate---}
As a first application, we consider achieving a non-linear phase shift to implement a control-Z gate between two photonic qubits.
Considering two photonic qubits, denoted $A$ and $B$ in dual-rail encoding $\ket{0}$ and $\ket{1}$, the control-Z gate realizes the transformation
\begin{align}
    \ket{x}_A \ket{y}_B \rightarrow (-1)^{xy} \ket{x}_A \ket{y}_B,
\end{align}
where $x$ and $y$ can take the value 0 or 1.
To implement this transformation, we consider the gate setup illustrated in Fig.~\ref{fig:CZ_gate}, inspired by Ref.~\cite{PhysRevA.95.062304}. 
First, the modes $\ket{1}_A$ and $\ket{1}_B$ are incident on a $50/50$ beam-splitter, which results in bunching due to the Hong-Ou-Mandel effect~\cite{HOM}, whereas other mode combinations (e.g., $\ket{0}_A\ket{1}_B$, etc.) travel through the circuit as independent photons.
Second, the photons are sent through the cascaded scattering and harmonic oscillator evolution as shown in Fig.~\ref{fig:intro-figure}b), where the $\ket{1}_A\ket{1}_B$ will pick up a phase $\phi_{NL}$ and the other combinations a phase $\phi_L$.
The cascading process is repeated until $\phi_{NL} - 2\phi_L = \pi$, and then the bunching of modes $\ket{1}_A$ and $\ket{1}_B$ is reversed by another $50/50$ beam splitter which finalizes the control-Z operation.
To characterize the performance of the control-Z gate, we use the Choi-Jamiolkowsky fidelity~\cite{JAMIOLKOWSKI, CHOI}, see the supplementary material for more information.
We numerically calculate the fidelity by propagating the wavefunctions in momentum space and vary it as a function of the number of scatterings up to $N=17$.
This is only done for an odd number of scatterings, as even numbers were found to yield poor fidelities.
We optimize the bandwidth $\sigma$, the detuning $\Delta$ between the central frequency of the pulses and the two-level emitter, and the harmonic trap parameters for each scattering number.
In Fig.~\ref{fig:CZ_gate_results}a), we show the optimized fidelities.
A fidelity of $\mathcal{F} \sim 75 \%$ is possible for $N=17$ scatterings without the trap, whereas with the trap, we reach a fidelity as high as $\mathcal{F} \sim 99.2 \%$.
Already at $N=3$ scatterings, the harmonic trap implementation reaches a fidelity $\sim 91\%$, which is higher than the best fidelity without a trap, and at $N=5$ scatterings, we reach a fidelity of $\sim  96\%$.
For the same fidelities we achieve with the harmonic trap, the implementation in Ref.~\cite{ShapiroCascadedCZGate} requires orders of magnitude more scatterings.
The implementation in Ref.~\cite{Schrinski_2022} requires the same number of two-level systems as we require scatterings, but is dependent on a specific identical level structure for all emitters. 
Overall, we see that the width of the pulse is almost constant (Fig.~\ref{fig:CZ_gate_results} b) and the parameters of the temporal trapping are almost unchanged as we vary $N$ (Fig.~\ref{fig:CZ_gate_results} d). 
The reason for the former is that it is advantageous to keep the photons within a time window of $\sim 1/\Gamma$ such that they interact with the emitter at the same time.
While the temporal trapping parameters are almost unchanged, the detuning increases with $N$ (Fig.~\ref{fig:CZ_gate_results} c).
As the detuning increases, the effective interaction strength between the photon decreases, allowing the temporal trap to better confine the system to the ground state for large $N$, even with fixed trapping parameters.
\begin{figure*}
\includegraphics[scale=0.55]{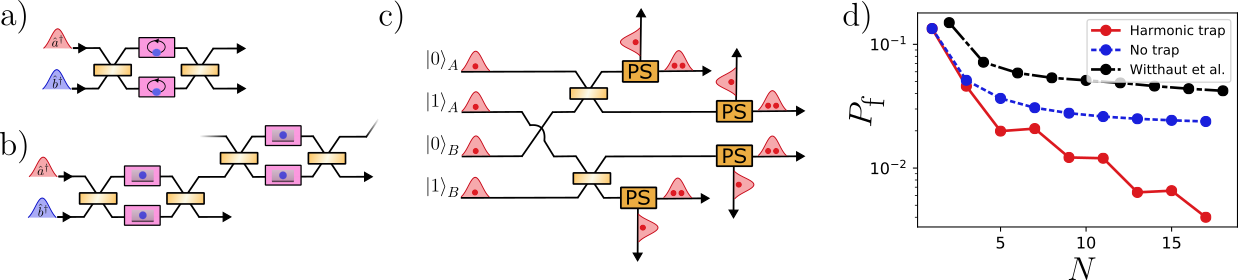}
\caption{a) The photon sorting device. The photon(s) enter the $\hat{a}$-mode and split on a $50/50$ beam splitter. They then enter the cascade scattering plus harmonic trap loop and leave after $N$-rounds. After the final beam splitter, if a one-photon pulse is sent in, it will leave through the $\hat{a}$-mode again; however, if a two-photon pulse is sent in, the photons will leave through the $\hat{b}$-mode. b) Photon sorting device without a temporal trap, the photon(s) scatter only once between the beam splitters. After one scattering, the sorting is not perfect, and thus, if the photon(s) leave the $\hat{a}$ mode, we scatter once more. This can then be cascaded $N$-times. c) Illustration of the Bell-state analyzer circuit, $PS$ stands for photon sorter. d) Failure probability $P_f$ as a function of the number of scatterings, with the photon sorter using a harmonic trap in a) (red), the photon sorter without a trap in b) (blue), and the proposal in Ref of Witthaut et al.~\cite{AndersPhotonSorter} (black).
 \label{fig:photon_sorter} }
\end{figure*}

\textit{Bell-state analyzer and photon sorting---}
A Bell-state analyzer is one of the fundamental building blocks to realize measurement-based~\cite{Oneway} and fusion-based quantum computing~\cite{FBQC} or building quantum repeaters~\cite{AllPhotonicRepeaters, SophiaRGS, LukinTwoWay, NicolasTwoWay}.
The Bell-state analyzer is a circuit that distinguishes the four Bell states $\ket{\phi^{\pm}} = \frac{1}{\sqrt{2}}(\hat{a}^{\dagger}_{0_A}\hat{a}^{\dagger}_{0_B} \pm \hat{a}^{\dagger}_{1_A}\hat{a}^{\dagger}_{1_B})\ket{\O}$, and $\ket{\psi^{\pm}} = \frac{1}{\sqrt{2}}(\hat{a}^{\dagger}_{0_A}\hat{a}^{\dagger}_{1_B} \pm \hat{a}^{\dagger}_{1_A}\hat{a}^{\dagger}_{0_B})\ket{\O}$,
where $A$ and $B$ refer to two different photonic qubits, and $0$ and $1$ represents the qubit encoding.
A linear-optical Bell-state analyzer can discriminate only two out of the four Bell states and thus succeeds with $p_s=50 \%$~\cite{browne2005, Gimeno2016, fusionrules}.
As an alternative approach, we present in the following a scheme to utilize the non-linear phase shift protocol to implement a Bell-state analyzer.
This involves using the non-linear phase shift to implement a photon sorter, and allows deterministic discrimination between all four Bell states, and is inspired by Ref.~\cite{AndersPhotonSorter}.
The photon sorter device is illustrated in Fig.~\ref{fig:photon_sorter}, and it is comprised of $50/50$ beam splitters and our cascaded scattering device, where Fig.~\ref{fig:photon_sorter}a) is the setup with a harmonic trap and Fig.~\ref{fig:photon_sorter}b) is the setup without it.
An incoming single-photon pulse entering through the $\hat{a}$-mode, will always leave the device through the $\hat{a}$-mode.
However, a two-photon pulse entering the $\hat{a}$-mode, will access the non-linearity from the cascading device and if a $\pi$ non-linear phase shift is attained without distorting the two-photon wavepacket away from the single-photon wavepacket, i.e., $\Psi(k,p) = -\Phi(k)\Phi(p)$, the photons will leave through the $\hat{b}$-mode.
That is, the device performs photon sorting by sending single-photon pulses to the output $\hat{a}$-mode and two-photon pulses to the output $\hat{b}$-mode.
For more details on the photon sorting device, see Ref.~\cite{AndersPhotonSorter} or the supplementary material.
With the photon sorting device, we can construct a Bell-state analyzer as illustrated in Fig.~\ref{fig:photon_sorter}c).
First, the $0$ and $1$ modes of the two photonic qubits are incident on a $50/50$ beam splitter.
The beam splitter bunches the photons of the $\ket{\phi^{\pm}}$ states in one of the modes, whereas for $\ket{\psi^{\pm}}$ the photons will be located in two different modes.
The photons are then sent through the photon sorting device, after which the $\ket{\psi^{\pm}}$ states are located in the $\hat{a}$-modes and the $\ket{\phi^{\pm}}$ states are located in the $\hat{b}$-modes with high probability.
With the $\ket{\phi^{\pm}}$ and $\ket{\psi^{\pm}}$ Bell states sorted, it is possible to discriminate the four Bell states through simple linear optics and standard photodetectors~\cite{AndersPhotonSorter}.
Importantly, within the approximation that we consider here, the setup never produces incorrect results.
Imperfect gate operation only results in inconclusive results, such that the bell state analyzer has a finite success probability $P_s = \frac{3}{4} - \frac{1}{4}\mathfrak{R} [\int dkdp \Phi^*(k)\Phi^*(p)\Psi(k,p)]$.
This is an important feature as heralded gate failure errors are easier to correct for than non-heralded gate errors in fault-tolerant photonic quantum computing~\cite{FBQC} and quantum communication~\cite{AllPhotonicRepeaters, pettersson2025QR}.
In Fig.~\ref{fig:photon_sorter}d), we plot the failure probability ($P_{\text{f}} = 1 - P_s$) as a function of the number of scatterings.
We plot our scheme using a harmonic trap in red, and without the trap in blue, and for comparison, we plot the results using the setup from Ref.~\cite{AndersPhotonSorter} in black.
Here, as for the control-Z gate, we see that using a harmonic trap results in significantly smaller failure probabilities, reaching a $P_f \sim 0.4\%$ with $N=17$ scatterings.
Moreover, already at $N=5$ we see a clear advantage of using the harmonic trap, where we achieve a failure probability of $P_f \sim 2\%$, which is already smaller than the minimum failure probability without a trap at $N=17$ ($P_f \sim 2.4\%$).
This near-deterministic Bell-state analyzer would greatly reduce the resource costs for fault-tolerant photonic quantum computing~\cite{FBQC, SegoviaMultiplex, löbl2023losstolerant} or long-distance quantum communication~\cite{AllPhotonicRepeaters}.
The optimal parameters for the harmonic trap in Fig.~\ref{fig:photon_sorter} d) are the same as for the Control-Z gate.

\textit{Conclusion---}
We have presented a scheme for realizing a high-fidelity non-linear phase shift on an itinerant light pulse.
The scheme relies on the repeated scattering off a two-level quantum emitter embedded in a chiral or one-sided waveguide, with a harmonic trap realized by a second-order dispersive element and a temporally varying phase.
The scheme reuses the same two-level emitter many times, compared to previous approaches where many two-level systems in the same waveguide are needed~\cite{Schrinski_2022, passiveCphaseGate}.
We analyzed the performance of this scheme for a control-Z gate setup to create entangled photonic qubits and for a passive Bell-state analyzer using a photon-sorter device.
We reach $\mathcal{F} \sim 99.2 \%$ ($\mathcal{F} \sim 96 \%$) for the control-Z gate and $P_s \sim 99.6 \%$ ($P_s \sim 98\%$) success probability for the Bell-state analyzer after $N = 17$ ($N = 5$) scatterings, both operations are of fundamental importance in photonic quantum information processing.
For instance, this near-deterministic Bell state measurement would allow for an increased photon loss threshold and reduce the resource cost in photonic quantum computing~\cite{FBQC, SegoviaMultiplex, löbl2023losstolerant}, or increase the communication rate in quantum communication~\cite{AllPhotonicRepeaters, pettersson2025QR}.
Furthermore, the high-fidelity control-Z gate would drastically increase the capabilities for photonic quantum simulations~\cite{KasperQuantumSim}.
\textit{Acknowledgments---}
We are grateful to Stefano Paesani, Kasper Nielsen, and Peter Lodahl for fruitful discussions.
We are grateful for financial support from Danmarks Grundforskningsfond (DNRF 139, Hy-Q Center for Hybrid Quantum Networks). 
L.A.P. acknowledges support from Novo Nordisk Foundation (Challenge project “Solid-Q”).
V.R.C. acknowledges support from the Danish National Research Foundation through the Center of Excellence for Complex Quantum Systems (Grant agreement No. DNRF 152).

\section{Supplementary Material} \label{sec:supplementary}
\subsection{Master equation approach}
In this section, we describe an approach to calculating the output quantum state of an emitter irradiated by a one or two-photon quantum pulse. To describe an incoming quantum pulse, we use the method of virtual cavities \cite{short_kiilerich, long_kiilerich, interaction-picture, virtual-cavity-review}. This method models the interaction between a quantum state in a pulse mode $u(t)$ and a quantum system irradiated by this quantum pulse by using a fictional (virtual) cavity with a time-dependent coupling coefficient designed to emit a pulse in the desired pulse shape. The time-dependent coupling coefficient that emits a mode in the shape $u(t)$ is given by
\begin{align}
    g_u(t) = \frac{u^*(t)}{\sqrt{1 - \int_0^t dt' |u(t')|^2}}.
\end{align}
Using this virtual cavity, the Hamiltonian for the interaction between the incoming pulse and the quantum emitter is
\begin{align}
    \hat{H} = \Delta \hat{\sigma}_+ \hat{\sigma}_- + \frac{i}{2} \sqrt{\gamma} \left( g_u(t) \hat{a}_u^\dagger \hat{\sigma}_- - \text{H.c.} \right),
\end{align}
where $\hat{a}_u$ annihilates an excitation in the quantum pulse mode, $\hat{\sigma}_-$ is the lowering operator for the quantum emitter, $\gamma$ is the decay rate of the quantum emitter, and H.c. denotes the Hermitian conjugate. The dynamics is accompanied by an associated dissipation operator given by
\begin{align}
    \hat{L} = g_u^*(t) \hat{a}_u + \sqrt{\gamma}\hat{\sigma}_-.
\end{align}
The output quantum state for the two-photon input state will populate the two-momentum $\Psi(k, p)$ wave function. Using this master equation approach, we can compute the two-time Fourier transform of the wave function $\tilde{\Psi}(t_1, t_2)$ with the following method. We initialize the system in the quantum state $\ket{\psi_0} = (\ket{0,g} + \ket{2_u, g})/\sqrt{2}$, where $2_u$ denotes two photons in pulse mode $u(t)$ and $g$ denotes the ground state of the emitter. The initial density matrix is therefore $\rho_0 = \frac{1}{2}(\ket{0}\bra{0} + \ket{0}\bra{2_u} + \ket{2_u}\bra{0} + \ket{2_u}\bra{2_u}) \otimes\ket{g}\bra{g}$. We can then evolve the initial quantum state from time $t_0$ to time $t_1$ with the master equation
\begin{align}
    \dot{\rho} = -i [\hat{H}, \rho] + \mathcal{D}[\hat{L}]\rho,
\end{align}
where $\mathcal{D}[\hat{L}]\rho = \hat{L}\rho \hat{L}^\dagger - \frac{1}{2}(\hat{L}^\dagger \hat{L} \rho  + \rho \hat{L}^\dagger \hat{L})$. At this point we apply the dissipation operator $\hat{L}(t_1)$ to the quantum state $\rho$ from the left as $\hat{L}(t_1) \rho$. We then integrate the master equation from $t_1$ to $t_2$ where we apply $\hat{L}(t_2)$ from the left as well, and trace over the result. Since we only apply $\hat{L}$'s from the left, only the $\ket{2_u}\bra{0}$ part of the initial state contributes to the trace, and since the destruction of the two photons in the $\ket{2_u}$-state happened at $t_1$ and $t_2$ respectively, the expectation value (the trace) is therefore equivalent to the two-time wave function,
\begin{align}
    \tilde{\Psi}(t_1, t_2) = \sqrt{2} \expval{\hat{L}(t_1)\hat{L}(t_2)},
\end{align}
where the factor $\sqrt{2}$ comes from the fact that the initial state carries a factor of $1/2$ on the part contributing to the expectation value, and applying $\hat{L}$ twice introduces a factor of $\sqrt{2}$. The method of virtual cavities only allows for input states in pulses which factors into single-time pulses, i.e. $\psi(t_1, t_2) = u(t_1)u(t_2)$. To scatter the output quantum state $\tilde{\Psi}(t_1, t_2)$ on the quantum emitter again, we need to perform a mode decomposition. Due to the bosonic symmetry of the photons, the Takagi-decomposition can exactly decompose the output quantum state into this form, when the quantum state is exactly populated by two photons \cite{Stolyarov-takagi, Paskauskas-takagi, FanPhotonSorter, photon-splitting},
\begin{align}
    \tilde{\Psi}(t_1, t_2) = \sum_i \lambda_i v_i(t_1) v_i(t_2).
\end{align}
Having performed this decomposition, we can iteratively scatter each mode $v_i(t)$ on the emitter, keeping in mind the relative population $\lambda_i$ of the mode.

\subsection{Trapping coefficients}
\label{supp:HarmoicOscDerivation}
In this section, we derive the coefficient to make the product of a temporally varying phase and a second-order dispersion time-evolution operator into an effective harmonic oscillator.
This derivation is done in position $x$ and momentum $p$ space, and afterwards we make the translation to time $t$ and frequency $\omega$ space.
The time evolution of the position operator $\hat{x}$ and momentum operator $\hat{p}$ under a harmonic oscillator Hamiltonian is given by~\footnote{With $\hbar = 1$.}:
\begin{align}
    \hat{x}(\Delta t) = \hat{x}(0) \cos{\left( \Omega \Delta t \right)} + \hat{p}(0)\sin{\left(\Omega \Delta t\right)}\sigma^2, \\
    \hat{p}(\Delta t) = \frac{-\sin{\left(  \Omega \Delta t \right)}}{\sigma^2} \hat{x}(0) + \cos{\left( \Omega \Delta t \right)} \hat{p}(0),
\end{align}
where $\sigma$ is the spatial width of a Gaussian. 
We wish to make time evolution of $\hat{x}$ and $\hat{p}$ under $U(t) = e^{-i\lambda_1 \hat{p}^2}e^{-i\lambda_2 \hat{x}^2}e^{-i\lambda_3 \hat{p}^2}$, to match the above equations. Leveraging the Baker-Hausdorff lemma \cite{BakerHausdorffLemma} we find:
\begin{align*}
    &\hat{x}(\Delta t) = \hat{x}(0)(1 - 4\lambda_2\lambda_3) + 2(\lambda_3 + \lambda_1(1 - 4\lambda_2 \lambda_3))\hat{p}(0), \\
    &\hat{p}(\Delta t) = -2 \lambda_2 \hat{x}(0) + (1 - 4 \lambda_2 \lambda_1)\hat{p}(0),
\end{align*}
which leads to the set of coupled equations:
\begin{align}
    (1 - 4\lambda_2 \lambda_3) &= \cos{\left(\Omega \Delta t\right)} \\
    2(\lambda_3 + \lambda_1(1 - 4\lambda_2 \lambda_3)) &= \sin{\left(\Omega \Delta t\right)}\sigma^2  \\
    2 \lambda_2 &= \frac{\sin{\left(\Omega \Delta t\right)}}{\sigma^2} \\
    (1 - 4 \lambda_2 \lambda_1) &= \cos{\left(\Omega \Delta t\right)}
\end{align}
This system can be solved with the following coefficients
\begin{align}
    \lambda_1 &= \tan{ \left(\frac{\Omega \Delta t}{2}\right)}\sigma^2/2, \\
    \lambda_2 &= \frac{\sin{\left(\Omega \Delta t\right)}}{2\sigma^2}, \\
    \lambda_3 &= \lambda_1,
\end{align}
From here, we can move to time and frequency space by the identification $x \xrightarrow{} t$ and $p \xrightarrow{} \omega$, i.e., position becomes time and momentum becomes frequency.
Thus, to realize the harmonic trap, we need to evolve the light pulses under $e^{-i\lambda_1 \omega^2}e^{-i\lambda_2 t^2}e^{-i\lambda_3 \omega^2}$. 
More concretely, experimentally, the harmonic trap is realized by: 
\begin{enumerate}
    \item First, applying a frequency-dependent phase shift with magnitude $\lambda_1$
    \item Second, we apply a time-dependent phase shift with magnitude $\lambda_2$.
    \item In the final step, we repeat step 1) again.
\end{enumerate}
That is, first we send the light pulses through a second-order dispersive element with a dispersive coefficient $\lambda_1$, then through a modulator with a temporal phase $\lambda_2$, and finally through a second-order dispersive element with a dispersive coefficient $\lambda_3$.
\subsection{Cancelling linear dispersion and absorbing second-order dispersion}
Scattering off the emitter introduces dispersion up to $n$-th order, which can be seen from the exponentiation of the linear transmission coefficient. That is $\mathcal{T}(k) \xrightarrow{} e^{\log(\mathcal{T}(k))}$ and expanding to $n$-th order. 
Expanding $\log{(\mathcal{T}(k))}$ to second order we get $e^{\log(\mathcal{T}(k))} \xrightarrow{} e^{\alpha_1 k + \alpha_2 k^2}$, with 
\begin{eqnarray}
    \alpha_1 = \frac{4 i \Gamma}{(2 \Delta_0-i \Gamma ) (2 \Delta_0+i \Gamma )} \\
    \alpha_2 = \frac{16 i \Gamma \Delta_0}{(2 \Delta_0-i \Gamma )^2 (2 \Delta_0+i \Gamma )^2},
\end{eqnarray}
where $\Delta_0$ is the detuning between the incoming pulse's central frequency and the two-level emitter's transition frequency.
The linear dispersion (i.e., $\alpha_1$) can be taken care of by a time delay.
Furthermore, the second-order dispersion (i.e., $\alpha_2$) can be absorbed in constructing the harmonic oscillator Hamiltonian.
\subsection{Choi-Jamiolkowsky fidelity}
Choi-Jamiolkowsky fidelity considers two Bell states as the input:
\begin{equation}
    \ket{\Psi_{\text{in}}} = \frac{1}{2}(\ket{00}_{AA'} + \ket{11}_{A A'}) \otimes (\ket{00}_{B B'} + \ket{11}_{B B'}), 
\end{equation}
where the gate only affects qubits $A$ and $B$. 
The fidelity is found by the overlap of the ideal state and the output state, $\mathcal{F} = |\bra{\psi_{\text{ideal}}}\ket{\psi_{\text{output}}}|^2$.
We use Gaussians $\phi(k) = e^{-(k-k_0)^2/2\sigma_{k}^2}/\sqrt{\pi^{1/2}\sigma_{k}}$ with spectral width $\sigma_k$ and central wave-number $k_0$ as input photon mode functions, and then use the linearly scattered photonic modes as the reference ideal mode function.
In this case, the fidelity can be calculated as
\begin{eqnarray}
    \mathcal{F} = \left| \frac{3}{4} -\frac{1}{4} \int dk dp \Phi^*(k)\Phi^*(p) \Psi(k, p)\right|^2,
\end{eqnarray}
where the ideal wavefunction $\Phi(k)\Phi(p)$ is that of two independent photons after the cascading, and $\Psi(k, p)$ is the wavefunction of the two photons after passing the non-linearity.
\subsection{Photon sorter details}
\label{supp:PhotonSorterDetails}
The incoming single-photon and two-photon pulses enter the photon sorter through the $\hat{a}$-mode.
A single photon pulse will split at the beam splitter $\hat{a}^{\dagger}_k \xrightarrow[]{} \left(\hat{a}^{\dagger}_k + \hat{b}^{\dagger}_k \right) / \sqrt{2}$, enter the cascaded scattering device where the $\hat{a}$ and $\hat{b}$ modes transform identically and then interfere at the beam splitter exiting in the $\hat{a}$ mode.
However, for a two-photon pulse, the beam-splitter will mix the $\hat{a}$ and $\hat{b}$ modes as $\hat{a}^{\dagger}_k\hat{a}^{\dagger}_p \xrightarrow[]{} \hat{a}^{\dagger}_k\hat{a}^{\dagger}_p + 2\hat{a}^{\dagger}_k\hat{b}^{\dagger}_p + \hat{b}^{\dagger}_k\hat{b}^{\dagger}_p$.
When the photons are in the same mode, they access the non-linearity from the cascading device. 
In contrast, if they are in separate modes, they will not access the non-linearity, as they enter the cascading device at different times.
After transforming under the cascaded scattering device and the final beam splitter, the output two-photon state is
\begin{eqnarray*}
    \ket{\Psi_{\text{out}}} = \frac{1}{2\sqrt{2}}\int dk dp ((\Phi(k)\Phi(p) + \Psi(k, p))\hat{a}^{\dagger}(k)\hat{a}^{\dagger}(p) + \\
    (-\Phi(k)\Phi(p) + \Psi(k, p))\hat{b}^{\dagger}(k)\hat{b}^{\dagger}(p))\ket{\O}.
\end{eqnarray*}
Again, here $\Phi(k)\Phi(p)$ is the transformed wavefunction of two independent photons, and $\Psi(k, p)$ is the transformed wavefunction of two photons accessing the non-linearity.
We see that if a $\pi$ non-linear phase shift has been attained without distorting the two-photon wavepacket away from the single-photon wavepacket, i.e., $\Psi(k,p) = -\Phi(k)\Phi(p)$, the photons will leave through the $\hat{b}$-mode.

\bibliography{bibliography}

\end{document}